# Wide-Range Bolometer with RF Readout TES

Sergey Shitov, Nikolay Abramov, Artem Kuzmin, Michael Merker, Matthias Arndt, Stefan Wuensch, Konstantin Ilin, Eugene Erhan, Alexey Ustinov and Michael Siegel

***Abstract*—To improve both scalability and noise-filtering capability of a Transition-Edge Sensor (TES), a new concept of a thin-film detector is suggested, which is based on embedding a microbridge TES into a high-Q planar GHz-range resonator weakly coupled to a 50-Ohm-readout transmission line. Such a TES element is designed as a hot-electron microbolometer coupled to a THz range antenna and as a load of the resonator at the same time. A weak THz signal coupled to the antenna heats the microbridge TES, thus reducing the quality factor of the resonator and leading to a power increment in the readout line. The power-to-power conversion gain, an essential figure of merit, is estimated to be above 10. To demonstrate the basic concept, we fabricated and tested a few submicron sized devices from Nb thin films for operation temperature about 5 K. The dc and rf characterization of the new device is made at a resonator frequency about 5.8 GHz. A low-noise HEMT amplifier is used in our TES experiments without the need for a SQUID readout. The optical sensitivity to blackbody radiation within the frequency band 600-700 GHz is measured as (2.7±0.9)×10$^{-14}$ W/√Hz at $T_c \approx$ 5 K at bath temperature ≈ 1.5 K.**

## I. Introduction

IMAGING applications in radio astronomy require large arrays of independent detectors (N ≥ 1000 pixels), which are cooled down to mK temperatures [1]. This raises issues with large heat load and demands reduction of wiring complexity to the cryogenic interface. A proven solution is the frequency-division multiplexing (FDM) [2] that means connection to different pixels at their individual frequencies using only one common signal line. There are at least two types of superconducting detectors employing FDM: transition-edge sensors (TES) [3] and microwave kinetic-inductance detectors (MKID) [4]. The TES is known as the most sensitive device in a wide frequency range; it employs relatively low FDM frequencies, since the output signal is boosted by a SQUID amplifier, which is demonstrated at frequencies up to a few MHz. The readout of a MKID can operate at much higher frequencies, at least up to 10 GHz, using compact integrated resonators and semiconductor RF amplifiers [4]. We have suggested a new solution for detector arrays [5], [6]: a THz antenna-coupled HEB micrbolometer [7] being embedded into a compact GHz-range resonator and operating at the superconducting transition. The resonator is excited via a common transmission line, similarly to MKID, providing each pixel of the array with its particular frequency slot. The response of a particular microbridge to the incident THz radiation appears in the common RF feed line as amplitude increment at its particular frequency. We have named this device RFTES. Since the output impedance is 50 Ohm, the response can be boosted by a semiconductor RF amplifier instead of an array of a few SQUIDs [1], [3]. Only two coaxial cables are needed for biasing and reading out such arrays. Since Q ≈ 10$^4$ is measured in our preliminary experiments, the frequency slots can be set by design as close as few MHz to each other. This approach may lead to kilo-pixel imaging arrays employing the new RFTES bolometers. It is worth to note here that we anticipate RFTES working like a resistive temperature detector being not limited to a particular frequency band and suitable to receive either centimeter or near IR waves, according to bandwidth of their antenna. Here, we present results of our study of single-pixel RFTES devices at signal frequency 645 GHz and probe frequency at about 5.8 GHz.

This work was supported in parts by grant 12-02-01352-a from Russian Foundation for Basic Research, by Ministry for Education and Science of Russian Federation with contracts 11.G34.31.0062 and K2-2014-025 (Program for Increase Competitiveness of NUST«MISiS»), and from German Federal Ministry of Education and Research 05K13VK4, 13N12025.

Sergey V. Shitov is with Kotel'nikov Institute of Radio Engineering and Electronics, Russian Academy of Sciences, Moscow 125009, Russia and with National University of Science and Technology MISiS, Moscow 119049, Russia (e-mail: sergey3e@gmail.com).

Artem A. Kuzmin (e-mail: artem.kuzmin@kit.edu), Michael Merker, Matthias Arndt, Stefan Wünsch, K. S. Ilin and M. Siegel are with Institute of Micro- and Nano-Electronic Systems and DFG-Center for Functional Nanostructures (CFN), Karlsruhe Institute of Technology, D-76187 Karlsruhe, Germany (e-mail: michael.siegel@kit.edu).

Nikolay N. Abramov and Eugene V. Erhan are with National University of Science and Technology MISiS, Moscow 119049, Russia (e-mail: n-abram-n@yandex.ru)

Alexey V. Ustinov is with Institute of Physics and DFG-Center for Functional Nanostructures (CFN), Karlsruhe Institute of Technology, D-76128 Karlsruhe, Germany and with National University of Science and Technology MISiS, Moscow 119049, Russia (e-mail: alexey.ustinov@kit.edu).

## II. Experimental Details

### A. Design and Fabrication

To verify our new concept experimentally, we have designed a layout, which can be used with various sizes of the absorber and at various temperatures. The samples are single-pixel detectors with a resonator frequency about 5.8 GHz fabricated using thin-film sputtering technology on sapphire substrates (ε ≈ 10) as presented in Fig. 1. The folded quarter-wave coplanar waveguide (CPW) resonator ($Z_0 \approx$ 70 Ohm) is patterned from 200-nm-thick films of Nb and weakly (-11 dB) coupled at its shorted end to the 50-Ohm-CPW readout line (the throughput line). The microbridge is also deposited using magnetron-sputtering process; the film thickness is 15 nm of Nb. The area of the microbridge film has been retained at 2 squares and varied for different devices in the range from 1 µm (length) by 0.5 µm (width) down to 0.2 µm by 0.1 µm. The impedance of the microbridge at THz frequency cannot be measured directly, so it is assumed to be similar to a witness device at about its normal state resistance just above $T_c$ (6-30 Ohm). The microbridge is coupled to a double-slot antenna via a specially designed impedance transformer. Due to tolerance of both the impedance of the microbridge and the fabrication process, the safety bandwidth of the transformer is estimated as BW ≈ 50-100 GHz centered at 645 GHz (not measured accurately). The antenna is positioned within the resonator near its open end at the point of nominal embedding impedance $R_e \approx 1$ Ohm.

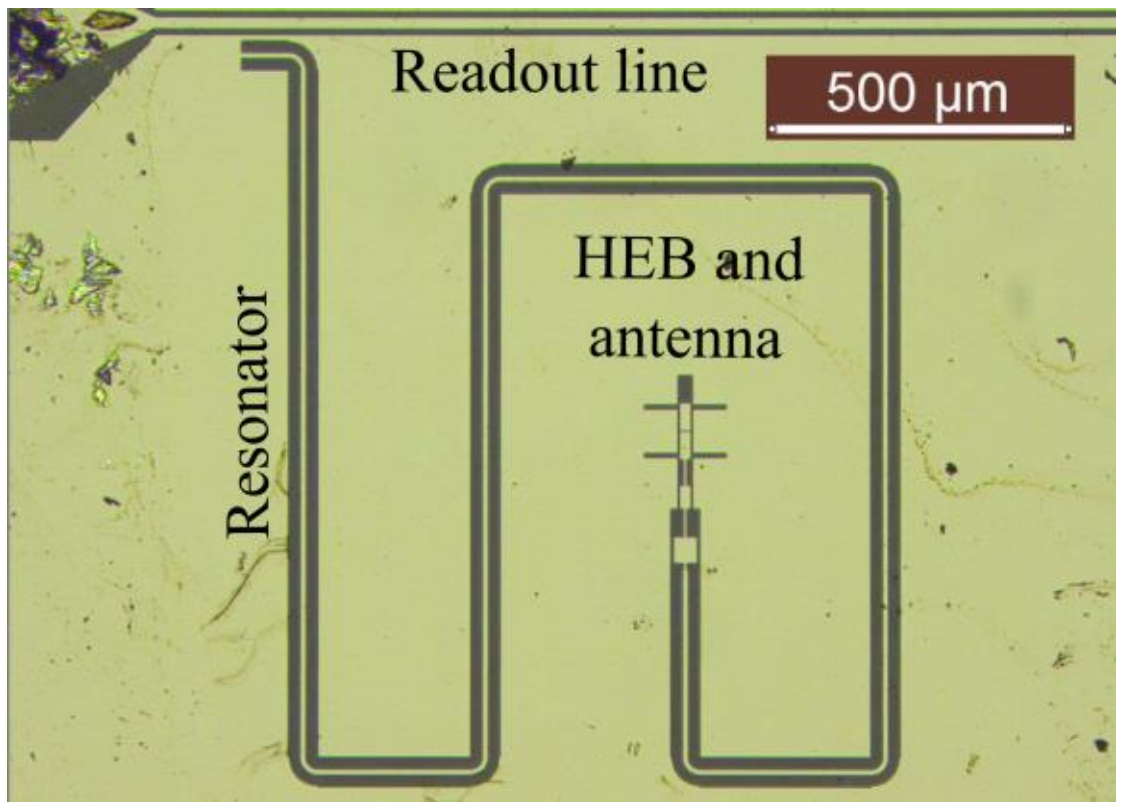

Fig. 1. Photo of the chip containing HEB microbridge integrated with 645-GHz double slot antenna and embedded into folded coplanar resonator at 5.8 GHz.

### B. *Experimental Setup*

Our cryostat [8] is equipped with a close-cycle pulse-tube refrigerator having separate stages at about 3 K and precooled $He^4$-Joule-Thomson stage suitable for temperatures down to 1.2 K. To perform optical measurements, the RFTES chip was integrated with a sapphire hemispherical immersion lens and thermally controlled via the temperature of the detector block, which cools the lens mount at the 1.2-K stage. The simplified configuration of the experiment is shown in Fig. 2(a). The more powerful 3-K stage cools the blackbody radiation source and the HEMT low-noise amplifier (LNA). The blackbody is designed similar to [9] as a multi-reflection 0.5-mm thick Ecosorb CR-100 RF absorber deposited inside a conical holder made of copper. The temperature of the BB is controlled via feedback loop, which includes a heater and a precision thermometer. The blackbody is placed inside the cryostat in front of the detector block, as shown in Fig. 2(a) to ensure that the main lobes of the antenna are completely covered by the aperture of the BB.

A vector network analyzer (Agilent, PNA-X series) was used as the probe source in our experiments. To prevent the room-temperature noise reaching the detector, the input path is equipped with a 3-K cooled 40-dB attenuator. The output path does not contain an RF isolator; the 20-cm coaxial cable is connecting the chip to the low-noise amplifier ($T_n \approx 10$ K,

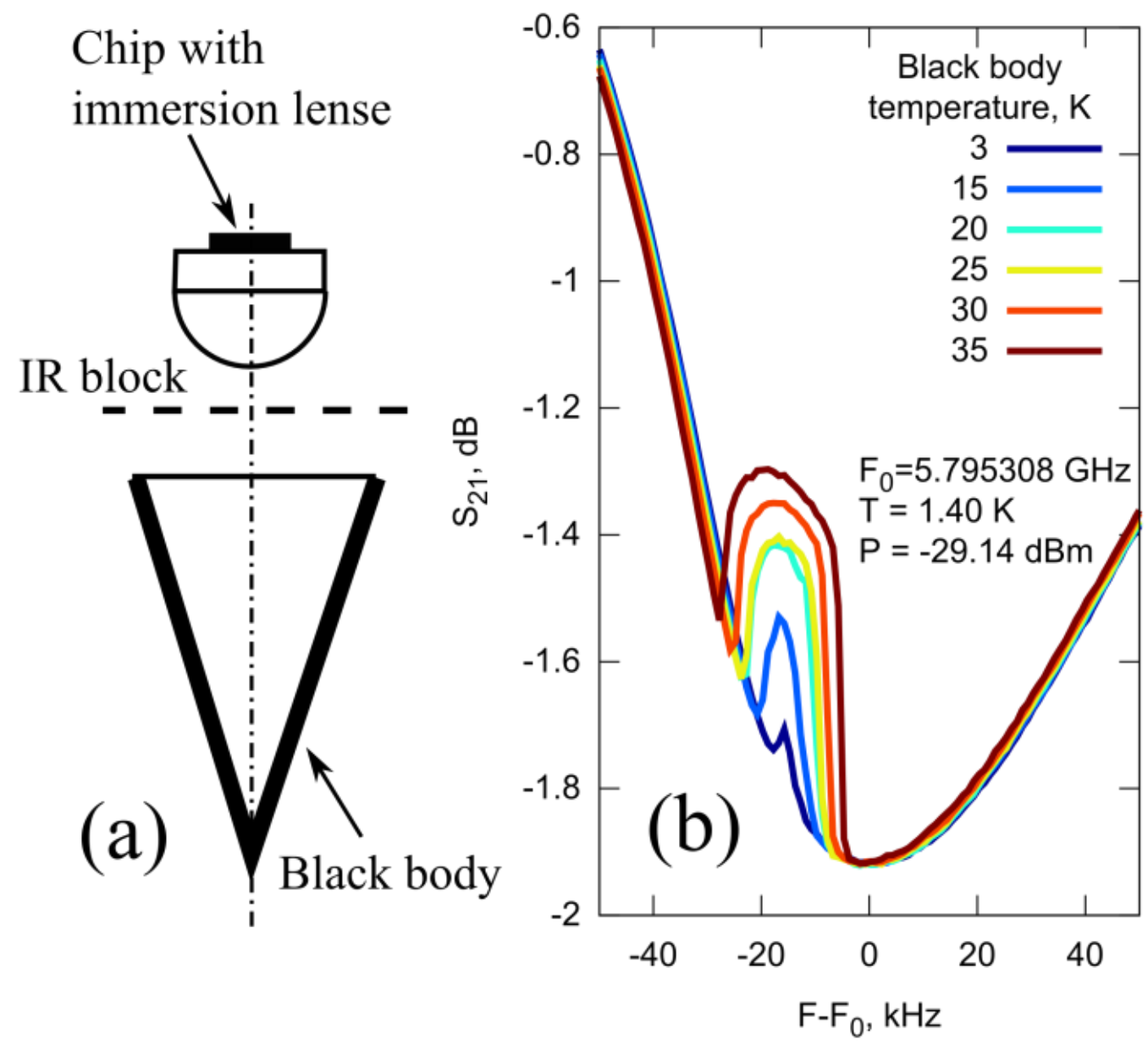

Fig. 2. Item (a) is a configuration of the experiment with blackbody radiation source. The radiation is low-pass filtered and coupled to the microbridge using integrated immersion lens antenna. Item (b) shows a response to the radiation near the resonance frequency $F_0$ at constant probe (bias) power -29.14 dBm applied to the throughput (readout) line; scattering parameter $S_{21}$ is normalized throughout the paper to a reference level -20.7 dB. Detector temperature is 1.4 K

G ≈ 26 dB) installed at the 3-K stage. We have also used a 6-dB attenuator at the output that is needed to prevent instability of the LNA. This combination unfortunately provides a noise figure of the output path (referred to the chip) of about 40-50 K.

### C. *Experimental data and Discussion*

The data from Fig. 2(b) depict the optical response of the experimental detector illuminated by radiation from the blackbody at different physical temperatures of the radiation source. We found that the transmission coefficient $S_{21}$ changes near the resonant frequency, if the temperature of the blackbody is changed. The operating point (the temperature of the electron gas) at the transition edge of the HEB is set via adjustment of the probe (bias) power at about 5.8 GHz. The response $S_{21}$ versus blackbody temperature (radiation temperature) has been measured at a fixed bias frequency as shown in Fig. 3. The bias frequency is adjusted to the center of the "inversed crater" from Fig. 2(b).

Since the measured signal frequency reaches THz range, and the physical temperature of blackbody is essentially low, the incident optical power must be calculated using Plank's formula, which accounts for both quantum noise effects and for the match of apertures between the antenna beam and the blackbody radiation source:

$$P(T) = \int_0^\infty \frac{2hf}{c^2} \frac{Tr(f)}{e^{hf/kT}+1} \cos(\theta) A\Omega df \qquad (1)$$

Here, *Tr*(*f*) is frequency-dependent transmission, *c* is speed of light, A is input aperture, Ω is solid angle. For our case of diffraction limited antenna beam ($A\Omega=\lambda^2$), which matches the aperture of blackbody by design, the formula (1) can be re-written as following:

$$P(T)=\int_{f_1}^{f_2}\frac{2h}{f}\frac{1}{e^{hf/kT}+1}df \qquad (2).$$

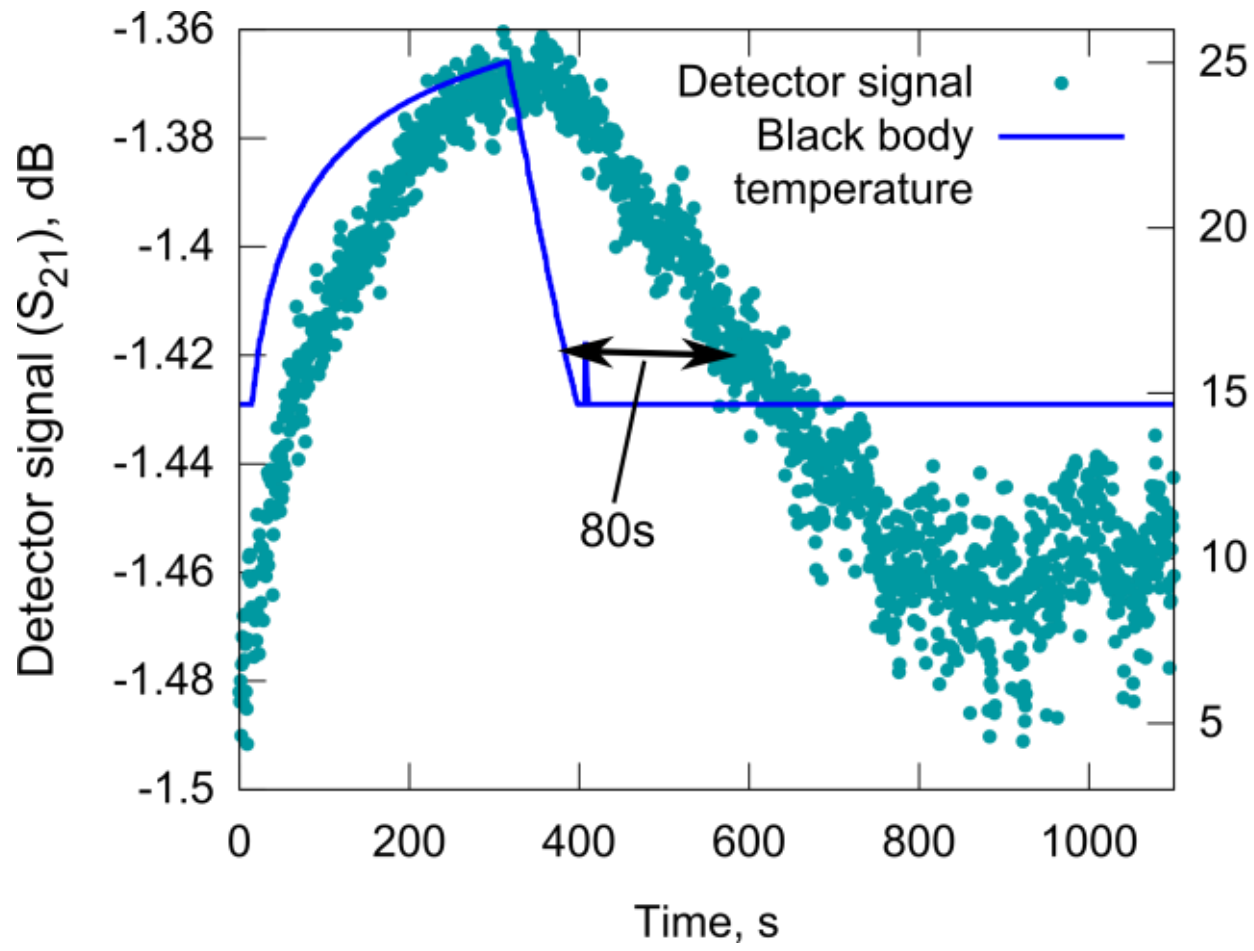

Fig. 3. Detector response $S_{21}$ and blackbody temperature on measurement interval of 1100 seconds. The arrow bar (not to scale) indicates the response lag due to low thermal conductivity of the CR-110 absorbing layer. Some drift can be noticed after $10^3$ seconds of experiment. Detector temperature is 1.4 K.

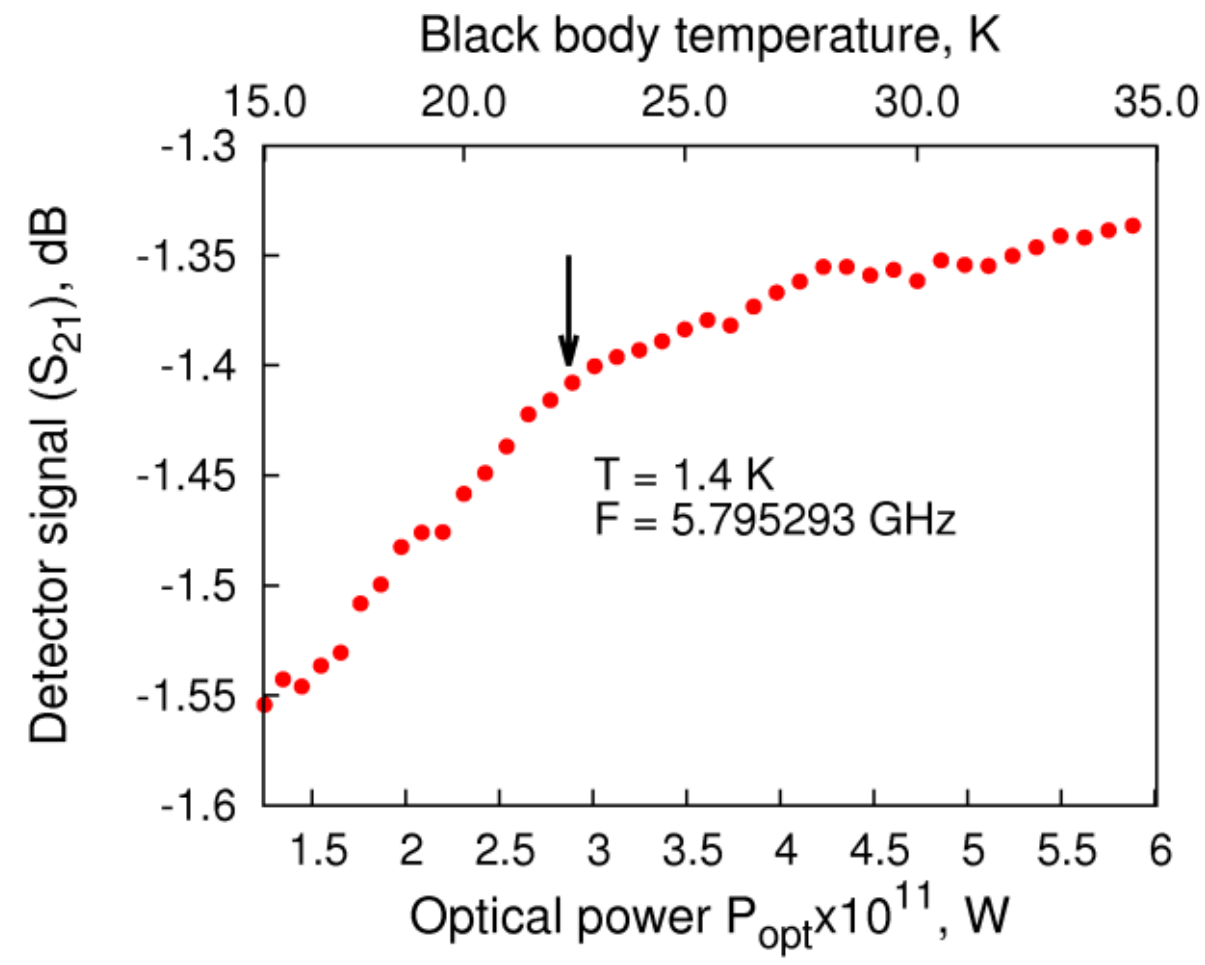

Fig. 4. Detector response $S_{21}$ on optical power and blackbody temperature. Some non-linear response might be suggested above the point marked with arrow. This effect can be, however, explained with change of feedback sign due to growing impedance of the microbridge.

It is possible to calculate from the experimental data the optical responsivity of the device, $S_{21}(P_{opt})$, as presented in Fig. 4, and to convert the measured noise spectrum into optical NEP as shown in Fig. 5, taking into account an antenna bandwidth ≈ 100 GHz centered at 645 GHz. It can be seen from Fig. 5 that the particular noise spectrum contains a 1/*f* component along with the white-noise component and with the commonly accompanying 100 Hz and 150 Hz interference. It is possible to analyze the dependence of the NEP spectrum to comply a general fitting expression: $S_{NEP}(f)=a+b/f$. Such form allows for separating the 1/*f* component from the white-noise components. The calculated spectral density of the white noise component yields an optical NEP ≈ $2.7\cdot10^{-14}$ W/√Hz. The detailed analysis of the noise will be discussed elsewhere. Nevertheless, it is possible to note right away that for $dS_{21}/df \neq 0$ (see Fig. 2) the phase noise of the VNA must be converted into the amplitude noise at the absorber. This effect is certainly harmful for a low-NEP operation of the bolometer.

Some other parasitic effects are studied. Since unwanted thermal links are possible within the restricted volume of the cryostat, and the heater of the BB is quite close to the detector block, it is important to be sure that there is no fake response in the output signal. In Fig. 6 (a) the response $S_{21}$ is measured versus probe (bias) power, while either BB temperature or bath temperature are changed. The first rather positive conclusion from the experimental data is that these variations have opposite signs. The increment of $S_{21}$ is negative for growing bath temperature and positive for higher temperature of the BB. In principle, the increment according to variations of the bath temperature has to be *positive* for a good device set to the very dip at $dS_{21}/df = 0$. The *negative* increment in $S_{21}$ presented on the bottom of Fig. 6 reflects the effect of growing kinetic inductance, which shifts the dip (the resonance frequency) to the lower frequency due to $dS_{21}/df < 0$. It was found experimentally that the presence of a parasitic impedance within the throughput line including its wire-bond connections can, according to our simulations, lead to asymmetry of the dip, thus resulting in the optimum operational point at $dS_{21}/df \neq 0$, as presented in Fig. 2. Since the condition $dS_{21}/df \neq 0$ is leading also to conversion of the phase noise to the amplitude noise of the probe signal, the possibility for improvements in the EM design and mounting procedure has to be studied in the nearest future.

To estimate the product of t optical coupling and power gain, $C_{OPT}G$, the concept of heat power substitution might be useful. The starting point can be the experimental fact of a systematic shift of the dependence of the probe power with varying the temperature of the BB that is clearly seen from Fig. 6 (a). Since the operation of the RFTES is suggested using fixed probe power, the effect can be registered as an increase of throughput power, while the temperature of the BB, $T_{BB}$, is increasing for a fixed probe power that is expressed as following,

$$\Delta P_{PROBE}=C_{OPT}G\Delta P_{OPT}(T_{BB}) \qquad (3).$$

Using data for $\Delta P_{OPT}$ from Fig. 4 and $\Delta P_{PROBE}$ from Fig. 6 (a) it is possible to estimate from (3) that the product $C_{OPT}G$ ranges from ≈ $10^3$ near the superconducting threshold at $S_{21}$ ≈ -22.3 dB to about $10^2$ at $S_{21}$ ≈ -21.9 dB. Because of the large power gain of the RFTES, the effect of the LNA on the performance of the system was hard to see in the experiment. It seems that the bolometer was operated in a regime with *positive* thermal electrical feed back, but still within the region of stability, which can be described with the following expression reflecting the increase of absorbed Joule power, $\Delta P_J$ with growing probe power due to *slow enough* growing of resistance of the microbridge, ΔR:

$$\Delta P_J - P_{PROBE}\Delta S_{31}(\Delta R) > 0 \quad (4)$$

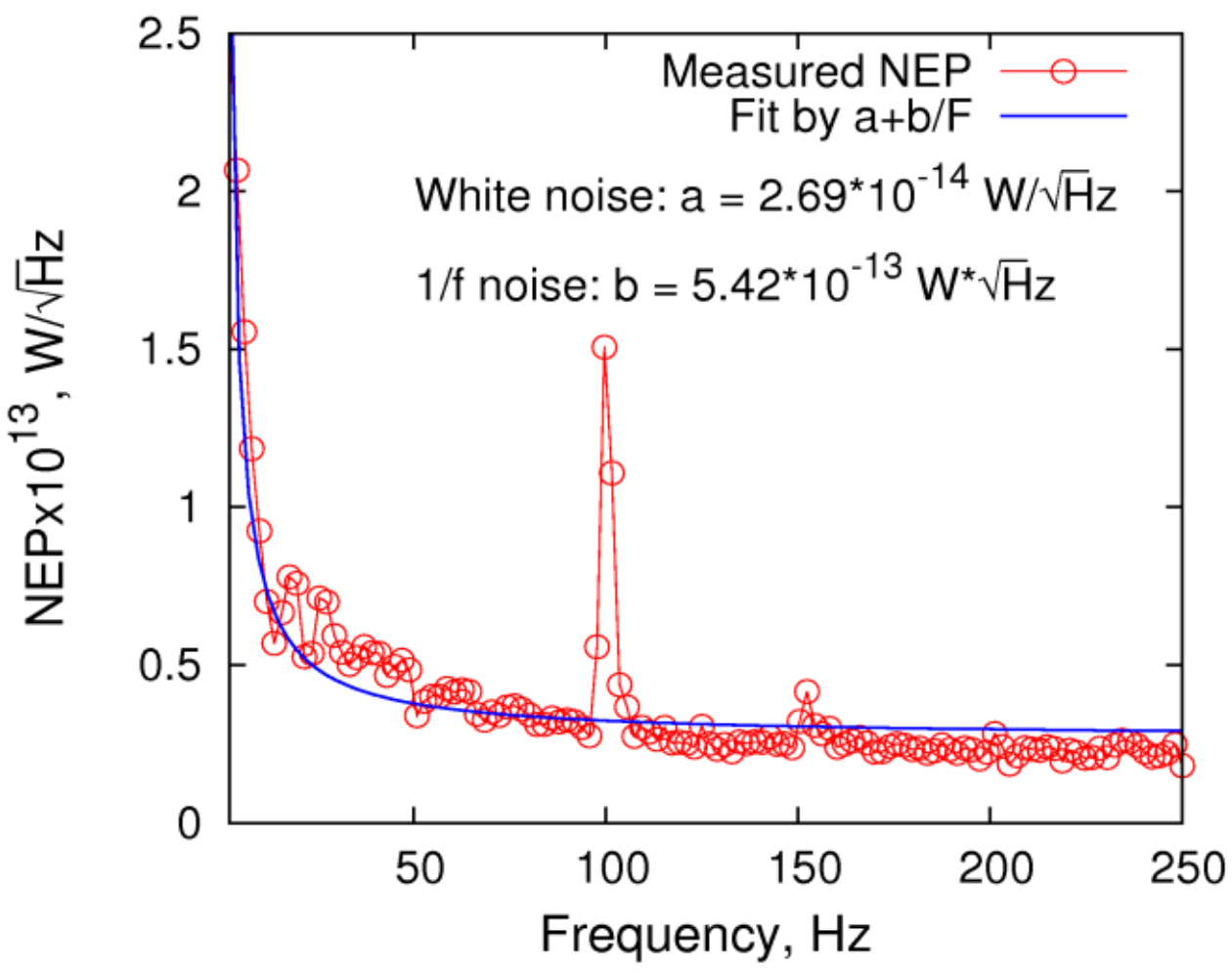

Fig. 5. Optical NEP calculated from experimental data for the optimal operating point (marked with arrow in Fig. 4).

This regime can be found at $R \le R_S$, i. e. near the matching point of the microbridge to the impedance of the resonator, $R_S$. The regime is, according to (4), always stable at $R > R_S$ due to $\Delta S_{31} < 0$ that means the decrease in the coupling efficiency while shifting to the classical regime of a voltage-biased TES bolometer.

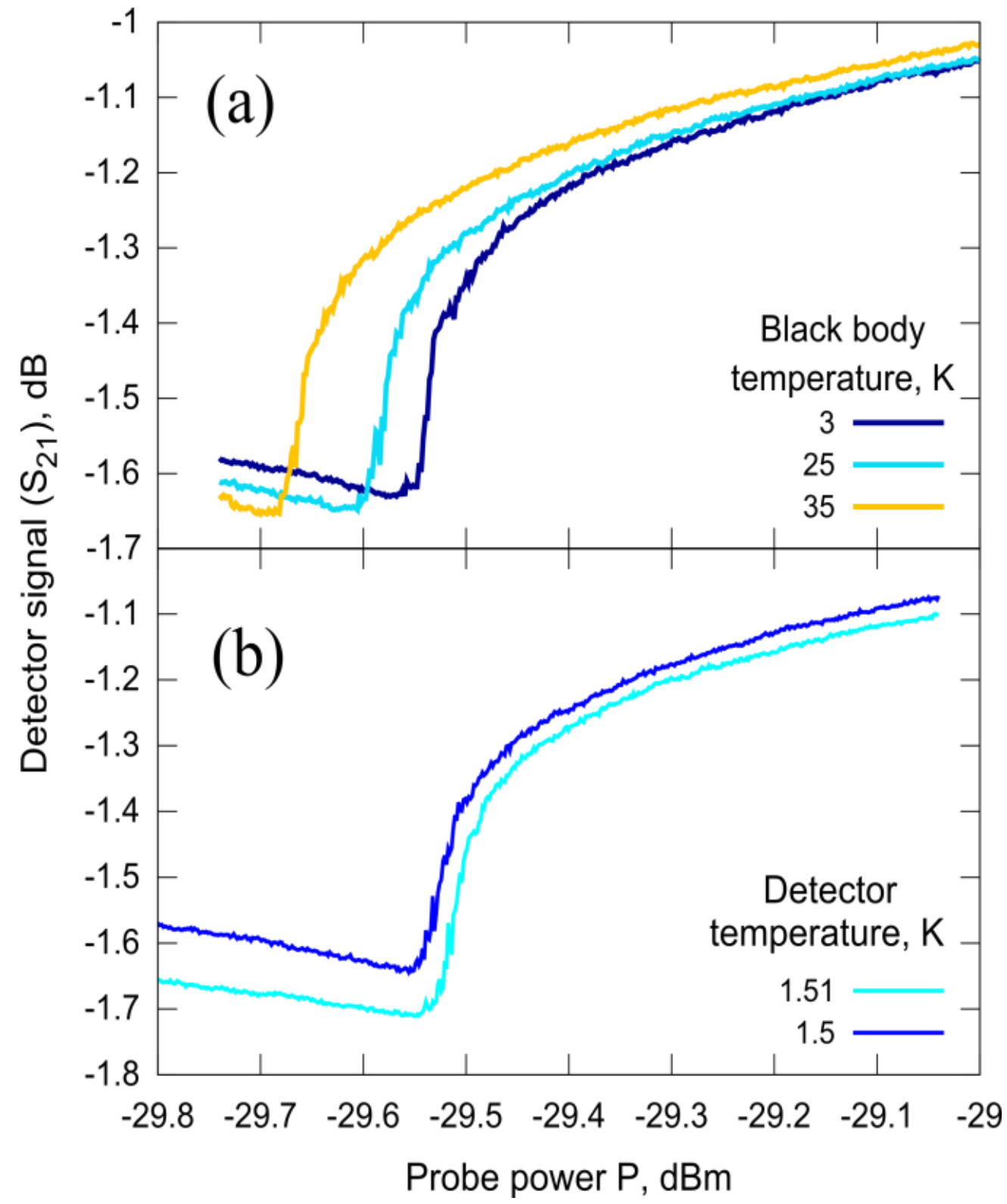

Fig. 6. Item (a) shows detector response $S_{21}$ versus bias power for a few blackbody temperatures. Detector temperature is 1.5 K. The experimental data in item (b) demonstrate that the detector optical response cannot be mixed up with (parasitic) heating of the detector block. The negative slope is the result of shift of resonance frequency due to small kinetic inductance effect in the superconducting microbridge. The positive slope is due to the growing RF loss at the superconducting transition.

## III. Conclusion

A new RFTES device is proposed, consisting of a hot-electron microbridge embedded in a microwave resonator, as an element of a sub-millimeter wave imaging array suitable for GHz-frequency FDM readout. The RFTES was fabricated and tested successfully demonstrating high power gain and good prospects of the new detector technology. The potential of the new device (RFTES) is clearly demonstrated with the total uncorrected optical NEP ≈ $(2.7\pm0.9)\cdot10^{-14}$ W/√Hz for a superconducting microbridge from Nb with $T_c \approx 5$ K operated at 1.5 K using a 10-K amplifier chain.

## Acknowledgment

The authors acknowledge fruitful discussions with Gregory Goltsman and Boris Karasik.

## References

[1] Wayne Holland, Michael MacIntosh, Alasdair Fairley et al., “SCUBA-2: a 10,000-pixel submillimeter camera for the James Clerk Maxwell Telescope,” in *Proc. of the SPIE*, vol. 6275, p. 62751E, 2006.

[2] T. M. Lantinga, Hsiao-Mei Choa, John Clarke, Matt Dobbs et al., “Frequency domain multiplexing for large-scale bolometer arrays,” *Millimeter and Submillimeter Detectors for Astronomy*, vol. 4855, pp. 172-181, 2003.

[3] K. D. Irwin, G. C. Hilton, “Transition-edge sensors,” *Topics in Applied Physics* vol. 99, pp.63-149, 2005.

[4] B. A. Mazin, “Microwave kinetic inductance detectors: The first decade,” in *Proc. of the LTD 13, AIP Conference Proceedings*, vol. 1185, pp. 135-142, 2009.

[5] S. V. Shitov, “Bolometer with high-frequency readout for array applications,” *Technical Physics Letters*, vol. 37, pp. 932–934, 2011.

[6] A. Kuzmin, S. V. Shitov, A. Scheuring, J. M. Meckbach, K. S. Il’in, S. Wuensch, A. V. Ustinov, M. Siegel, "Development of TES bolometers with high-frequency readout circuit," *IEEE Transactions on Terahertz Science and Technology*, 2013. DOI: 10.1109/TTHZ.2012.2236148 .

[7] B. S. Karasik, W. R. McGrath, M. E. Gershenson, A. V. Sergeev, "Photon-noise-limited direct detector based on disorder-controlled electron heating," *J. Appl. Phys.* **87**, 7586, 2000.

[8] Triton Helium-4 refrigerator - 1 K. Oxford Instruments, http://www.oxford-instruments.com/

[9] A. V. Uvarov, S. V. Shitov, A. N. Vystavkin, "A cryogenic quasioptical millimeter and submillimeter wavelength bands blackbody calibrator," *Measurement Techniques*, vol. 53, No. 9, pp. 1047-1054, 2010.